# Adaptive resolution for multiphase smoothed particle hydrodynamics

Xiufeng Yang, Song-Charng Kong*

Department of Mechanical Engineering, Iowa State University, Ames, IA 50011, USA

*Corresponding author: kong@iastate.edu

## Abstract

The smoothed particle hydrodynamics (SPH) method has been increasingly used to study fluid problems in recent years; but its computational cost can be high if high resolution is required. In this study, an adaptive resolution method based on SPH is developed for multiphase flow simulation. The numerical SPH particles are refined or coarsened as needed, depending on the distance to the interface. In developing the criteria, reference particle spacing is defined for each particle, and it changes dynamically with the location of the interface. A variable smoothing length is used together with adaptive resolution. An improved algorithm for calculating the variable smoothing length is further developed to reduce numerical errors. The proposed adaptive resolution method is validated by five examples involving liquid drops impact on dry or wet surfaces, water entry of a cylinder and dam break flow, with the consideration of ambient gas. Different resolution levels are used in the simulations. Numerical validations have proven that the present adaptive resolution method can accurately capture the dynamics of liquid-gas interface with low computational costs. The present adaptive method can be incorporated into other SPH-based methods for efficient fluid dynamics simulation.

# 1. Introduction

Smoothed particle hydrodynamics (SPH) is a Lagrangian particle method without the use of computational grids. Due to its advantages in simulating free surface flows, multiphase flows and moving boundaries, the SPH method has been applied to solve various fluid problems [1, 2]. In the literature, the simulations of incompressible flows using SPH are usually based on uniform spatial resolution, which may require a large number of numerical particles, incurring long computer time.

To reduce the computational demand, several methods based on non-uniform resolution have been developed for use in SPH simulations. In simulating the self-gravitating collapse, Kitsionas and Whitworth [3] increased the resolution in high-density regions by splitting particles. Feldman and Bonet [4] presented a particle splitting method for dynamic particle refinement and studied the numerical errors introduced by particle splitting. López et al. [5] presented a dynamic refinement algorithm with application to free-surface flows and non-cohesive soil problems. Vacondio et al. [6] developed a variable-resolution method, in which one parent particle was split into seven child particles for refinement, while two child particles were merged into one parent particle for coarsening. Barcarolo et al. [7] also developed particle refinement and coarsening techniques. During the refinement process, one parent particle was split into four child particles and the parent particle was turned off but not erased. After leaving the refinement region, the parent particle was turned on and the child particles were erased. However, it was found that the total mass did not always conserve during this process [7]. Khorasanizade and Sousa [8] presented a dynamic flow-based particle splitting method. Inspired by the idea of adaptive mesh refinement from the mesh-based method, Chiron et al. [9] developed an adaptive particle refinement technique for SPH, in which guard particles were used between adjacent domains with different resolutions. Later, Sun et al. [10] also applied the adaptive particle refinement technique in their work.

The previous methods on non-uniform resolution based on SPH only consider single-phase flows and are not fully "adaptive." In order to change the resolution, a particle splitting technique is usually used to refine particles, and a particle merging technique is needed to coarsen the particles. In addition to particle splitting and merging, the criteria for refinement and coarsening are required. In general, there are two types of refinement, namely, static and dynamic [5]. For static resolution, the refinement zones are static in space and known beforehand; all the particles that enter the refinement zones are split into smaller particles [4, 6, 7, 9, 10]. Since the refinement zone is static, the static resolution is not adaptive. In contrast, for dynamic resolution, the refinement zones move according to the physical properties of the flows, such as density [3], velocity [5], and vorticity [8]. The dynamic resolution method is partially adaptive because the refinement criteria based on density, velocity or vorticity need to be defined beforehand and are different for different cases.

In the present paper, we proposed a fully adaptive resolution method for SPH simulation of multiphase flows. The spatial resolution varies with the location of the interface between different phases. The dynamics of the interface does not need to be known beforehand.

## 2. SPH Method

The Naiver-Stokes (NS) equations are used to describe fluid flows:

$$\frac{d\rho}{dt} = -\rho \nabla \cdot \boldsymbol{u} \tag{1}$$

$$\frac{d\boldsymbol{u}}{dt} = \boldsymbol{g} - \frac{1}{\rho}\nabla p + \frac{\mu}{\rho}\nabla^2 \boldsymbol{u} \tag{2}$$

where $\rho$ is density, $\boldsymbol{u}$ is velocity, $p$ is pressure, $\mu$ is dynamic viscosity, and $\boldsymbol{g}$ is the gravitational acceleration.

In SPH, a function $f$ and its derivative $\nabla f$ can be calculated using particle summations:

$$f_i = \sum_j \frac{m_j}{\rho_j} f_j W(r_{ij}, h) \tag{3}$$

$$\nabla f_i = \sum_j \frac{m_j}{\rho_j} (f_j - f_i) \nabla_i W(r_{ij}, h) \tag{4}$$

where the subscripts $i$ and $j$ denote SPH particles, $m$ is particle mass, $W$ is kernel function, $r$ is the distance between two particles, and $h$ is smoothing length. To avoid the so-called tensile instability of SPH that may occur in fluid simulations, the kernel function suggested by Yang et al. [11, 12] is used:

$$W(s, h) = \frac{1}{3\pi h^2} \begin{cases} s^3 - 6s + 6, & 0 \leq s < 1 \\ (2-s)^3, & 1 \leq s < 2 \\ 0, & 2 \leq s \end{cases} \tag{5}$$

where $s = r/h$.

For adaptive resolution, the variable smoothing length should be used. In order to conserve momentum, the symmetric expressions of kernel function suggested by Hernquist and Katz [13] are employed:

$$\overline{W}_{ij} = \frac{1}{2} \left[ W(r_{ij}, h_i) + W(r_{ij}, h_j) \right] \tag{6}$$

$$\nabla_i \overline{W}_{ij} = \frac{1}{2} \left[ \nabla_i W(r_{ij}, h_i) + \nabla_i W(r_{ij}, h_j) \right] \tag{7}$$

In a number of previous studies [14, 15], the algorithm of variable smoothing length proposed by Hernquist and Katz [13] was used to update the smoothing length, which can be written in 2D as:

$$h_i^{n+1} = \frac{h_i^n}{2}\left(1 + \sqrt{\frac{N_r}{N_i^n}}\right) \tag{8}$$

where $n$ is timestep. $N_r$ is a reference number of neighbor particles. Since the standard SPH is based on constant smoothing length, the use of variable smoothing length can potentially introduce numerical errors [16]. To reduce the numerical errors caused by variable smoothing length, Eq. (8) is improved as follows.

$$h_i^{n+1} = \frac{1}{2}\left[\frac{h_i^n}{2}\left(1 + \sqrt{\frac{N_r}{N_i^n}}\right) + \frac{1}{N_i^n}\sum_j h_j^n\right] \tag{9}$$

The smoothing length of particle $i$ is determined by not only the number of its neighboring particles but also the average smoothing length of its neighbors. It should be noted that in our simulations, the smoothing length is updated every timestep, while in the literature [13-15] the smoothing length is usually updated when $N_i$ differs from $N_r$ by more than a prescribed tolerance. Because the variable smoothing length is used, the tree search algorithm [17] is used for searching neighboring particles.

In this work, the kernel width is expected to be $3\Delta s$, here $\Delta s$ is the particle spacing, then $N_r$ can be estimated as follows.

$$N_r = \frac{\pi(3\Delta s)^2}{\Delta s^2} \approx 28 \tag{10}$$

In SPH, the NS equations (1) and (2) can be replaced by the following particle equations:

$$\frac{d\rho_i}{dt} = \sum_j m_j(\boldsymbol{u}_i - \boldsymbol{u}_j) \cdot \nabla_i \overline{W}_{ij} \tag{11}$$

$$\frac{d\boldsymbol{u}_i}{dt} = \boldsymbol{g} - \sum_j m_j \left(\frac{p_i + p_j}{\rho_i \rho_j} + \Pi_{ij}\right)\nabla_i \overline{W}_{ij} + \sum_j \frac{4 m_j \mu_i \mu_j (\boldsymbol{r}_i - \boldsymbol{r}_j) \cdot \nabla_i \overline{W}_{ij}}{\rho_i \rho_j (\mu_i + \mu_j)(r_{ij}^2 + \eta)}(\boldsymbol{u}_i - \boldsymbol{u}_j) \tag{12}$$

The artificial viscosity proposed by Monaghan [1] is used.

$$\Pi_{ij} = \begin{cases} \dfrac{-\alpha(c_i + c_j)\mu_{ij} + 2\beta\mu_{ij}^2}{(\rho_i + \rho_j)}, & (\boldsymbol{u}_i - \boldsymbol{u}_j)\cdot(\boldsymbol{r}_i - \boldsymbol{r}_j) < 0 \\ 0, & (\boldsymbol{u}_i - \boldsymbol{u}_j)\cdot(\boldsymbol{r}_i - \boldsymbol{r}_j) \geq 0 \end{cases} \quad (13)$$

$$\mu_{ij} = \dfrac{(h_i + h_j)(\boldsymbol{u}_i - \boldsymbol{u}_j)\cdot(\boldsymbol{r}_i - \boldsymbol{r}_j)}{2(r_{ij}^2 + \eta)} \quad (14)$$

The parameters $\alpha$ and $\beta$ in Eq. (13) are used to control the strength of the artificial viscosity. In the present simulations, $\alpha = 0.01$ and $\beta = 0$ for the liquid phase [18], and $\alpha = 1$ and $\beta = 2$ for the gas phase [19]. The term $\eta = 0.01(h_i + h_j)^2/4$ is added to prevent the singularity when two particles are too close to each other [19].

In SPH, the simulation of incompressible flows usually uses uniform particle distribution. The particle splitting and/or merging processes may result in a non-uniform distribution, which will cause additional forces between the particles. In order to adjust the particle distribution, the particle shifting technique [20] was applied in a previous study [8]. However, the particle splitting technique may led to unexpected particle diffusion, especially when there is a free surface or interface [21]. Thus, in this paper a repulsive force is considered when two particles are too close to each other.

$$\dfrac{d\boldsymbol{u}_{ij}^c}{dt} = \chi\left(\dfrac{r_c}{r_{ij}}\right)^\varsigma m_j \dfrac{p_i + p_j + p_r}{\rho_i \rho_j} \nabla_i \overline{W}_{ij}, \quad r_{ij} < r_c \quad (15)$$

Here the superscript and subscript $c$ denotes the variables related to the two close particles. The parameters $\chi$, $\varsigma$ and $r_c$ are used to control the magnitude and shape of the repulsive force. In 2D space, $r_c \equiv \varepsilon(\sqrt{V_i} + \sqrt{V_j})/2$ and $\varepsilon < 1$, indicating that the repulsive force is only valid when the distance between two particles is less than the expected particle spacing. This force can also prevent particle penetration.

Pressure is calculated using the following equation of state, which is commonly used in SPH simulation,

$$p = c^2(\rho - \rho_r) + p_r \quad (16)$$

where $c$ is a numerical speed of sound, $\rho_r$ is a reference density, $p_r$ is a reference pressure.

To reduce the pressure fluctuation caused by density evaluation, density is reinitialized every 20 timesteps using the Shephard filter [22] as shown in Eq. (17). It should be noted that in Eq. (17) a cubic spline kernel function is used, as shown in Eq. (18). Eq. (18) will result in less numerical errors than Eq. (5) in evaluating density when Eq. (17) is used.

$$\bar{\rho}_i = \frac{\sum_j m_j W_{ij}}{\sum_j V_j W_{ij}} \tag{17}$$

$$W(s,h) = \frac{5}{14\pi h^2} \begin{cases} (2-s)^3 - 4(1-s)^3, & 0 \leq s < 1 \\ (2-s)^3, & 1 \leq s < 2 \\ 0, & 2 \leq s \end{cases} \tag{18}$$

## 3. Adaptive resolution method

The process to change the resolution adaptively is as follows. A reference particle spacing is defined for each particle, and it changes dynamically with the distance to the liquid-gas interface. With the reference particle spacing and reference density, a reference particle mass can be find. If the ratio of the mass of a particle to its reference mass is less than a given value used for particle splitting, the particle will be split into smaller particles; if the mass ratio is less than a given value used for particle merging, the particle will be merged to one of its neighboring particles. Therefore, the resolution is controlled by the reference particle spacing.

### 3.1 Reference particle spacing

The reference particle spacing of a particle is defined as a function of the distance between the particle and the interface of different phases. The reference particle spacing has a large value when the particle is far away from the interface. If the particle comes close to the interface, its reference particle spacing decreases.

To find the reference particle spacing quickly, particle band is used, as shown in Fig. 1. The width of particle band is $K\Delta S$, where $\Delta S$ denotes the reference particle spacing and $K$ is a parameter used to control the band width. For the same particle band, the particles have the same value of reference particle spacing. The particle band should not be too wide or too narrow. Considering the interaction between neighboring particles, $K\Delta S$ is set to be the width of the kernel function. The relationship between the reference particle spacing of adjacent particle bands is defined as

$$\Delta S_{k+1} = C_r \Delta S_k \tag{19}$$

where $k$ (= 0, 1, 2, …) is the index of particle band. The parameter $C_r$ is called adaptive number, which is the ratio of the reference particle spacing of two adjacent particle bands, and it controls the size of the refinement areas. When $C_r = 1$, the particle spacing is uniform, namely, the resolution does not change adaptively. When the value of $C_r$ increases, the particle spacing increases with the increase of distance from the interface. The value of $C_r$ is in the range of $1 \leq C_r \leq 1.2$.

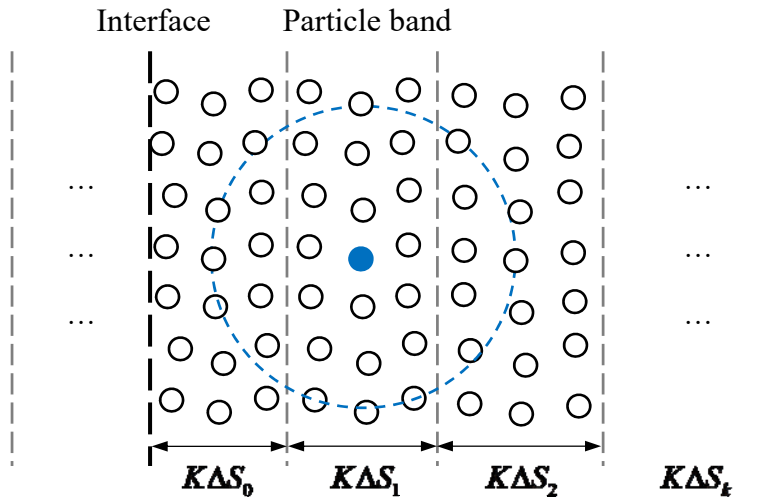

Fig. 1. Reference particle spacing and particle band

The tree search algorithm, used for searching neighboring particles, is also used to search the particles in particle bands. The reference particle spacing is updated every 200 timesteps.

## 3.2 Particle splitting and merging

In our previous work [23], an algorithm of particle splitting and merging was proposed to avoid a large particle mass difference caused by mass transfer near the liquid-gas interface due to evaporation. The previous algorithm for particle splitting and merging is improved for adaptive resolution in the present work. The major improvement is that in this work each particle has its own reference mass, while the reference mass is the same for all the particles of the same phase in the previous work [23].

The reference mass in the two-dimensional simulation is defined as

$$m_r = \rho_r (\Delta S)^2 \qquad (20)$$

If the ratio between the mass of a particle and its reference mass is larger than the number $\gamma_s$ for particle splitting, namely,

$$m/m_r > \gamma_s \qquad (21)$$

the particle will be split into two smaller particles, as shown in Fig. 2. If the mass ratio is less than the number $\gamma_m$ for particle merging, namely,

$$m/m_r < \gamma_m \qquad (22)$$

the particle will be merged with its nearest neighboring particle, as shown in Fig. 3.

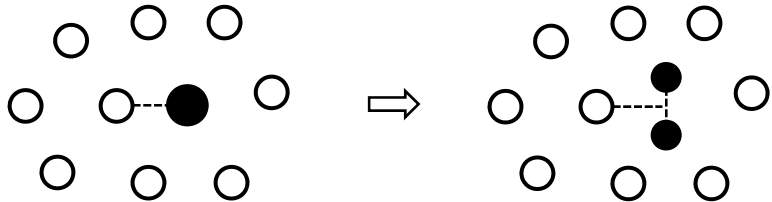

Fig. 2. Schematic of particle splitting

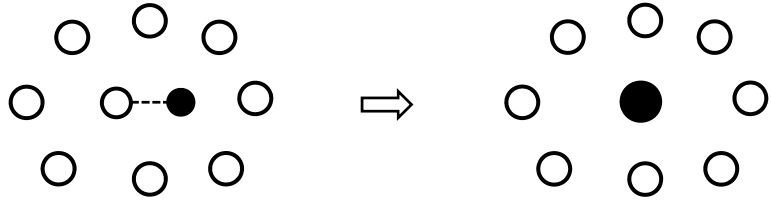

Fig. 3. Schematic of particle merging.

To find the proper values of $\gamma_s$ and $\gamma_m$, we assume

$$\begin{cases} m_s = 2m_m \\ \dfrac{m_s}{m_r} = \dfrac{m_r}{m_m} \end{cases} \quad (23)$$

where

$$\begin{cases} m_s = \gamma_s m_r \\ m_m = \gamma_m m_r \end{cases}. \quad (24)$$

The assumption of Eq. (23) is based on the idea that the mass ratios caused by particle splitting and merging should be the same. Substituting Eq. (24) in Eq. (23) and solving Eq. (25) will yield the following.

$$\begin{cases} \gamma_s = 2\gamma_m \\ \gamma_s = \dfrac{1}{\gamma_m} \end{cases} \quad (25)$$

$$\begin{cases} \gamma_s = \sqrt{2} \\ \gamma_m = \dfrac{\sqrt{2}}{2} \end{cases} \quad (26)$$

Considering that $\gamma_s = 2\gamma_m$ may lead to oscillation between particle splitting and merging, $\gamma_s$ should be larger than $2\gamma_m$. Thus, this study uses $\gamma_s = 1.5$, $\gamma_m = 0.7$.

The method to determine the particle spacing between the two child particles is as follows. If the particles are uniform before splitting, as shown in Fig. 4(a), it is straightforward to find a splitting pattern, after which the particle distribution is also uniform. In this case, the particle distance between the child particles is $\frac{\sqrt{2}}{2}\Delta s \approx 0.7\Delta s$. If the particles are split uniformly in a line, as shown in Fig. 4(b), the particle distance between the child particle is $0.5\Delta s$. Since these are two extreme cases, the particle spacing is generally in the range of $0.5\Delta s$ to $0.7\Delta s$.

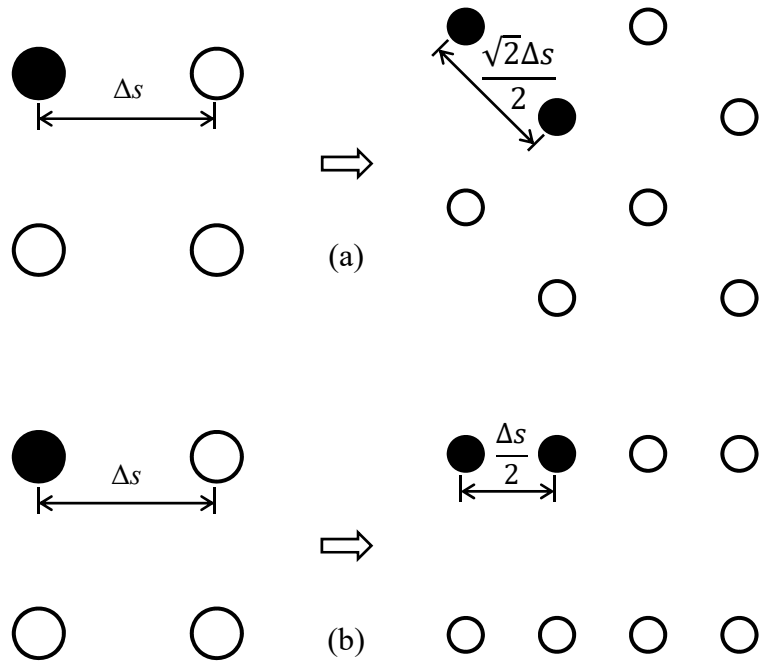

Fig. 4. (a) The particles are uniform before and after splitting. (b) The particles are uniform before splitting but uniformly split in a line.

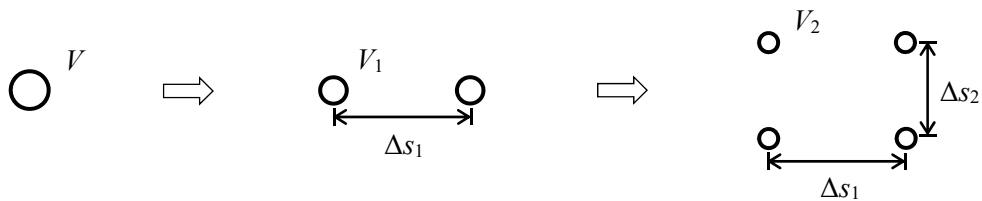

Fig. 5. One particle is split into two particles and then four particles

Fig. 5 shows that one particle is split into two particles, which are split into four particles subsequently. The ratio of "particle spacing between the child particles" to "particle spacing of between the parent particles" is kept the same. It is reasonable to assume that the volume does not change during the process of particle splitting. In 2D space, the following is obtained.

$$V = \Delta s^2, \ V_1 = \frac{V}{2} = \frac{\Delta s^2}{2}, \ V_2 = \frac{V_1}{2} = \frac{\Delta s^2}{4} \tag{27}$$

The value of $V_2$ can also be calculated as

$$V_2 = \Delta s_1 \Delta s_2 \tag{28}$$

The ratio of the particle spacing between the child particles to the particle spacing of the parent particles is denoted by $\lambda$.

$$\lambda = \frac{\Delta s_1}{\Delta s} = \frac{\Delta s_2}{\Delta s_1} \tag{29}$$

Solving Eqs. (27) and (28) yields $\lambda \approx 0.6$. This value is consistent with the range of $0.5 \leq \lambda \leq 0.7$, as indicated above.

## 4. Numerical examples

The present adaptive resolution method is tested by simulating five different cases, namely, 1) drop impact on a wet surface, 2) drop impact on a dry surface, 3) drop impact on a deep pool, 4) water entry of a horizontal cylinder, and 5) dam break flow. The first three cases are used to validate the adaptive resolution by comparison with the uniform resolution. The last two cases are used to validate the adaptive resolution by comparison with experiments from literature.

**4.1 Drop impact on a wet surface**

Fig. 6 shows the computational setting. Here $D = 1$ mm is the drop diameter, $H_d$ is the distance between the drop and the film, $H_f = D/4$ is film thickness, $L$ is the width of the computational domain, and $H_d = D/2$ is the distance from the liquid-gas interface to the top boundary of the computational domain. The computational domain is $20D$ in width and $10D$ in height. The bottom face is a solid surface, while the top face uses a mirror boundary condition. Periodic boundary conditions are used on both the left and right faces. The initial drop velocity $U_0$ is 5 m/s. The densities of the liquid and gas phases are 1,000 kg/m$^3$ and 1.29 kg/m$^3$, respectively. The dynamic viscosities of the liquid and gas are 0.001 kg/m/s and $2\times10^{-5}$ kg/m/s, respectively.

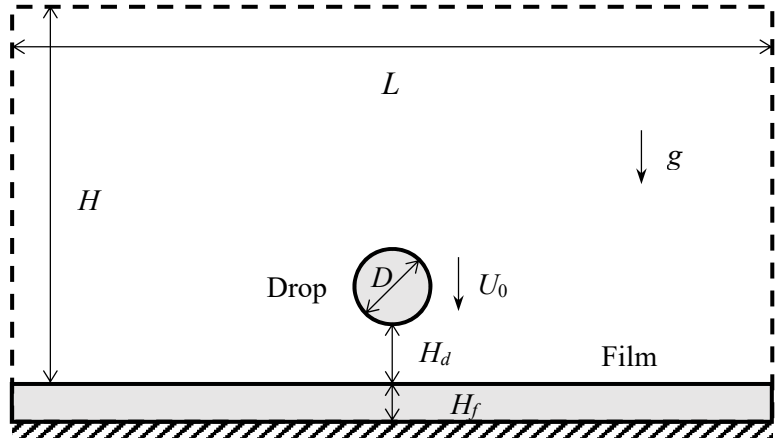

Fig. 6. Computational settings of drop impact on a surface

The particle spacing of different resolution levels is given in Table 1. The particle spacing varies from Level 3 to Level 0 for adaptive resolution. In the following simulations with adaptive resolution, the finest resolution is used at the liquid-gas interface, and the coarsest resolution is used in the region far away from the interface. Because the number of liquid particles is much less than the number of gas particles, the adaptive resolution is only used for the gas phase, while the liquid phase is simulated at uniform resolution Level 3. Simulations using uniform resolutions are also conducted for validating the present adaptive resolution methodology.

Table 1. Particle spacing of different resolution levels.

| Resolution | Level 0 | Level 1 | Level 2 | Level 3 | Adaptive |
|---|---|---|---|---|---|
| Particle spacing (mm) | 0.2 | 0.1 | 0.05 | 0.025 | 0.025 ~ 0.2 |

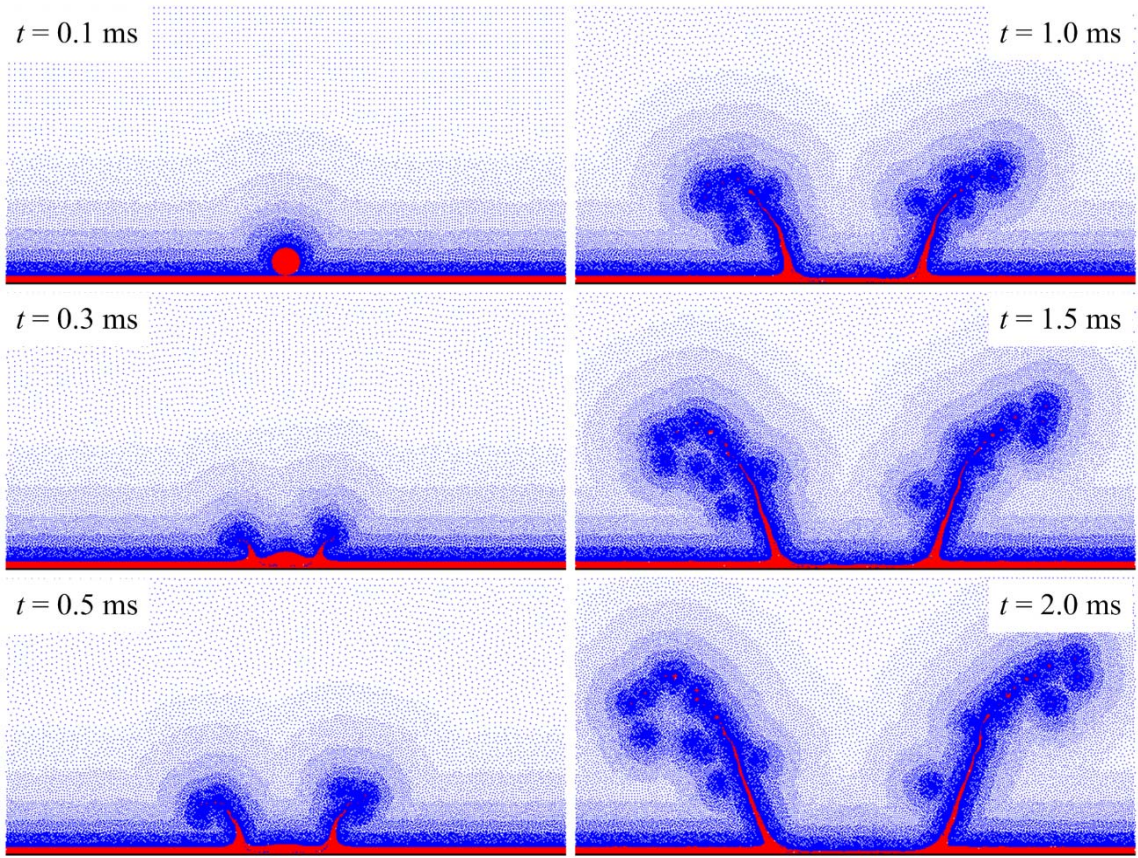

Fig. 7. History of drop impact on a wet surface simulated using adaptive resolution for $C_r = 1.1$

Fig. 7 shows the process of drop impact on a wet surface predicted by SPH simulation using adaptive resolution $C_r = 1.1$. It can be seen that a crown-like structure is formed when the drop impacts the film. The crown will fall down and finally merge into the liquid film. Since this work is focused on testing the adaptive resolution method, the simulation only runs to $t = 2$ ms. The resolution of the gas phase near the interface is at the finest level, while the resolution is gradually coarsened away from the interface. It can also be seen that the refinement region adaptively follows the evolution of the liquid-gas interface. Note that there is no need to know or define the refinement region beforehand. Thus, the present method is a true "adaptive" resolution method.

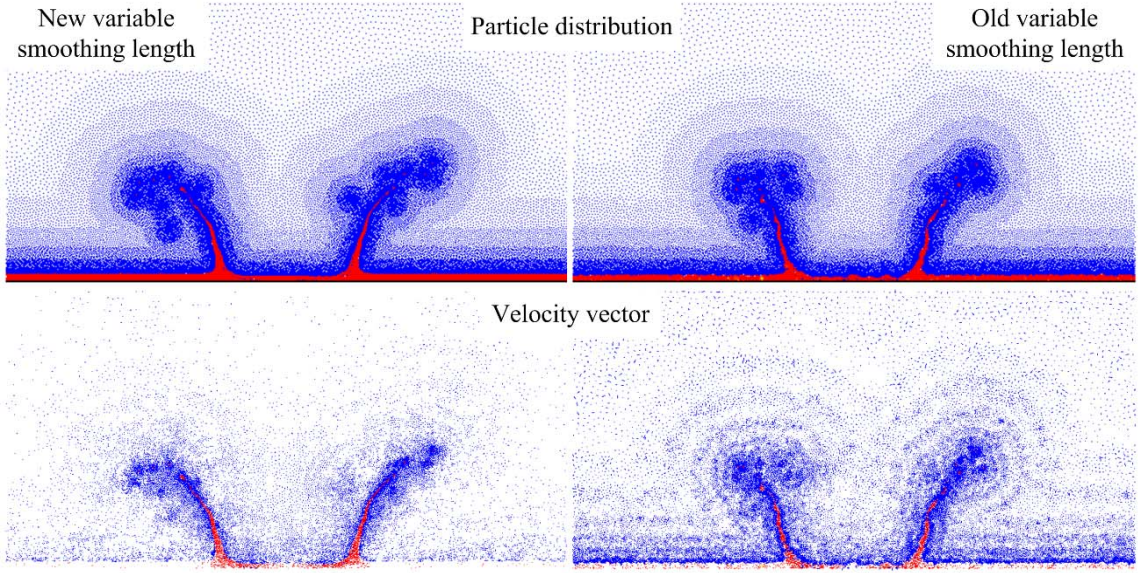

Fig. 8. Comparison of results at $t = 1$ ms for adaptive resolution $C_r = 1.1$ using the new variable smoothing length (Eq. (9)) and the old variable smoothing length (Eq. (8))

It should be noted that the numerical results in Fig. 7 are obtained using the improved variable smoothing length (Eq. (9)). As shown in Fig. 8, the predicted liquid-gas interface using Eq. (9) is much smoother than that using Eq. (8), since the new variable smoothing length (Eq. (9)) can reduce the numerical noise caused by the variable smoothing length Eq. (8). In fact, even for the simulation using constant resolution, Eq. (9) also has performed better than Eq. (8). Fig. 9 compares the particle distribution and velocity vector at $t = 2$ ms for simulations with resolution Level 1 using constant smoothing length (top), new variable smoothing length (middle) and old variable smoothing length (bottom).

Results using the new variable smoothing length agree better with those using constant smoothing length.

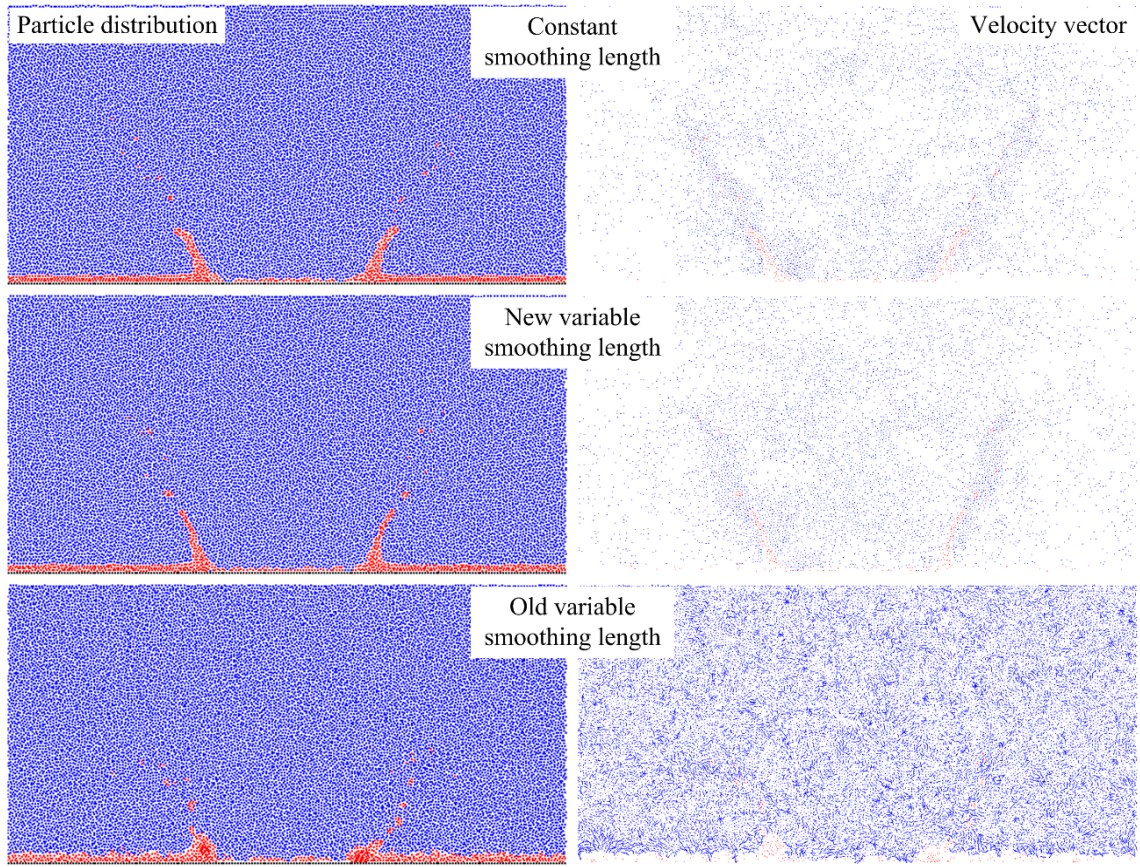

Fig. 9. Comparison of particle distribution and velocity vector at $t = 2$ ms from three different simulations using uniform resolution Level 1

Fig. 10 compares the simulation results of drop impact using uniform resolution with constant smoothing length and adaptive resolution with improved variable smoothing length. It can be seen that the crowns at different uniform resolutions have similar shapes with almost identical widths but very different heights. Compared to resolution Level 3, cases in Level 1 and Level 2 do not have enough particles to form fine crowns. The result from resolution Level 3 can be regarded as the baseline. The results from adaptive resolution $C_r = 1.05$, 1.1 and 1.2 agree well with the baseline, while they use much less gas particles.

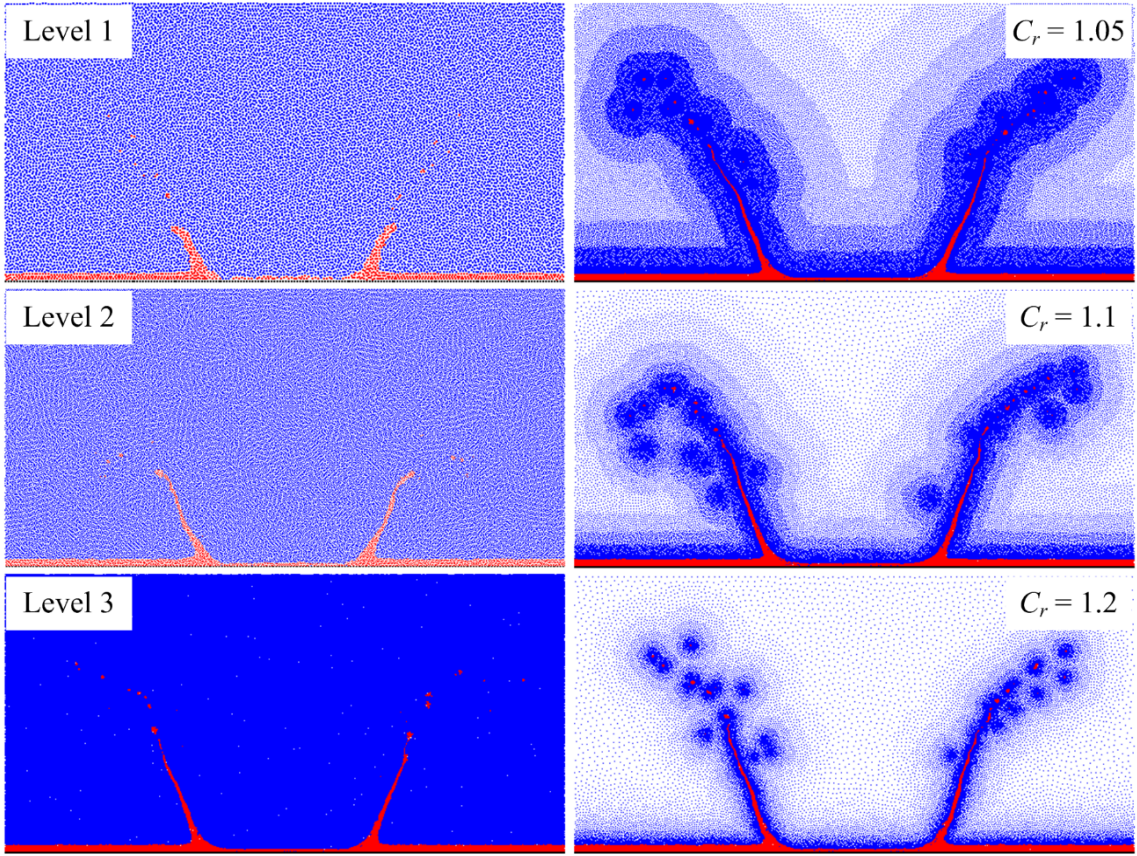

Fig. 10. Comparison of results at $t = 2$ ms using uniform resolutions and adaptive resolutions

To compare the number of particles with the finest resolution (Level 3), the relative number of particles ($R_N$) is defined as

$$R_N = \frac{N}{N_{L3}} \times 100\% \qquad (30)$$

where $N$ is the number of particles, and $N_{L3}$ is the number of particles for uniform resolution Level 3. Fig 11 shows the relative number of gas particles for different resolutions. The relative numbers of resolutions Level 1 and Level 2 are 6.25% and 25% of the baseline, respectively. The particle number for adaptive resolution is not constant, and it varies according to the shape of interface. The gas particles used for adaptive resolution $C_r = 1.05$ varies from 14% to 30% of the baseline, and the relative number of gas particles for $C_r$ = 1.1 and 1.2 are only 8~20% and 5~11%, respectively. The present adaptive resolution method has proven to significantly reduce the number of particles, hence the computational cost.

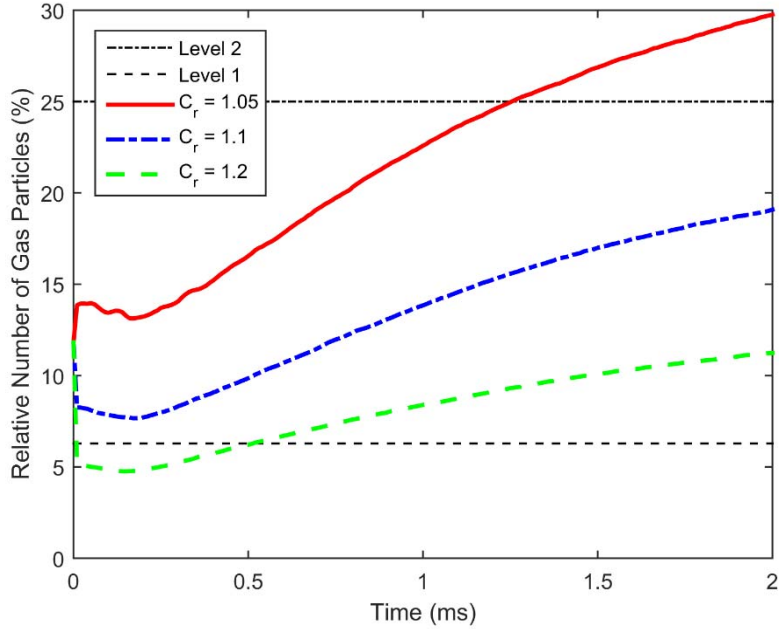

Fig. 11. Relative number of gas particles as a function of time for various cases. The relative particle number for resolution Level 3 is 100%.

Table 2 shows the computational time of the simulations using different resolutions. For uniform resolutions Level 1 to Level 3, the CPU time increases about 8 times when the particle spacing decreases 2 times (the particle number increases 4 times and the number of time step increases 2 times), which indicates that the CPU time is approximately proportional to the number of particles and the number of time steps. For adaptive resolution $C_r$ = 0.05, 0.1, and 0.2, the simulations are about 2.9, 4.3, and 6.1 times faster than the simulation of uniform resolution Level 3, respectively.

Table 2. Comparison of CPU time of different resolutions for drop impact on a wet surface.

| Resolution | Particle spacing (mm) | CPU time (minute) |
|---|---|---|
| Level 1 | 0.1 | 10 |
| Level 2 | 0.05 | 76 |
| Level 3 | 0.025 | 609 |
| $C_r = 0.05$ | 0.025 ~ 0.2 | 209 |
| $C_r = 0.1$ | 0.025 ~ 0.2 | 141 |
| $C_r = 0.2$ | 0.025 ~ 0.2 | 100 |

**4.2 Drop impact on a dry surface**

For the case of drop impact on a dry surface, the computational setting is similar to that of the wet surface case shown in Fig. 6, but there is no film on the solid surface. The distance between the drop and the solid wall $H_d$ is $D/2$. The computational domain is $20D$ in width and $10D$ in height. As shown in Fig. 12, when the drop impacts the dry wall, it spreads on the solid surface and forms a very thin film. Tiny droplets are generated beyond the edge of the film; the refinement region around the droplets can also be seen at $t = 1.0$ ms in Fig. 12.

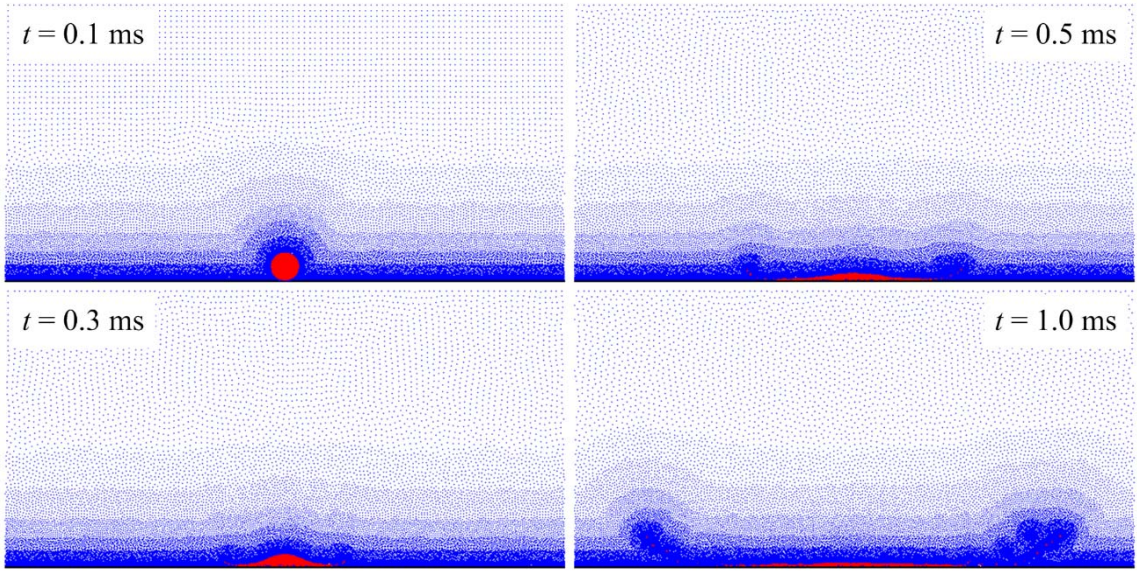

Fig. 12. Simulations of drop impact on a dry surface using adaptive resolution $C_r = 1.1$

The impact of three consecutive drops on a surface can result in more complex outcome. In this simulation, the initial distance between the centers of the adjacent drops is 2$D$. Fig. 13 shows the simulation results using adaptive resolution $C_r$ = 1.1. Similar to the single drop case, the first drop forms a film on the dry wall. Then the second drop impinges on the film formed by the first drop. Similar to the case of drop impact on a wet surface in Section 4.1, the second drop also forms a crown. However, the film formed by the first drop is much thinner and smaller than the film in Section 4.1, thus the shape of the crown formed by the impingement of the second drop is not the same as the crown formed in Section 4.1. The impact of the third drop produces an additional crown (Fig. 13 at $t$ = 1 ms and 1.5 ms). In the simulation, the refinement regions move adaptively with the liquid-gas interfaces.

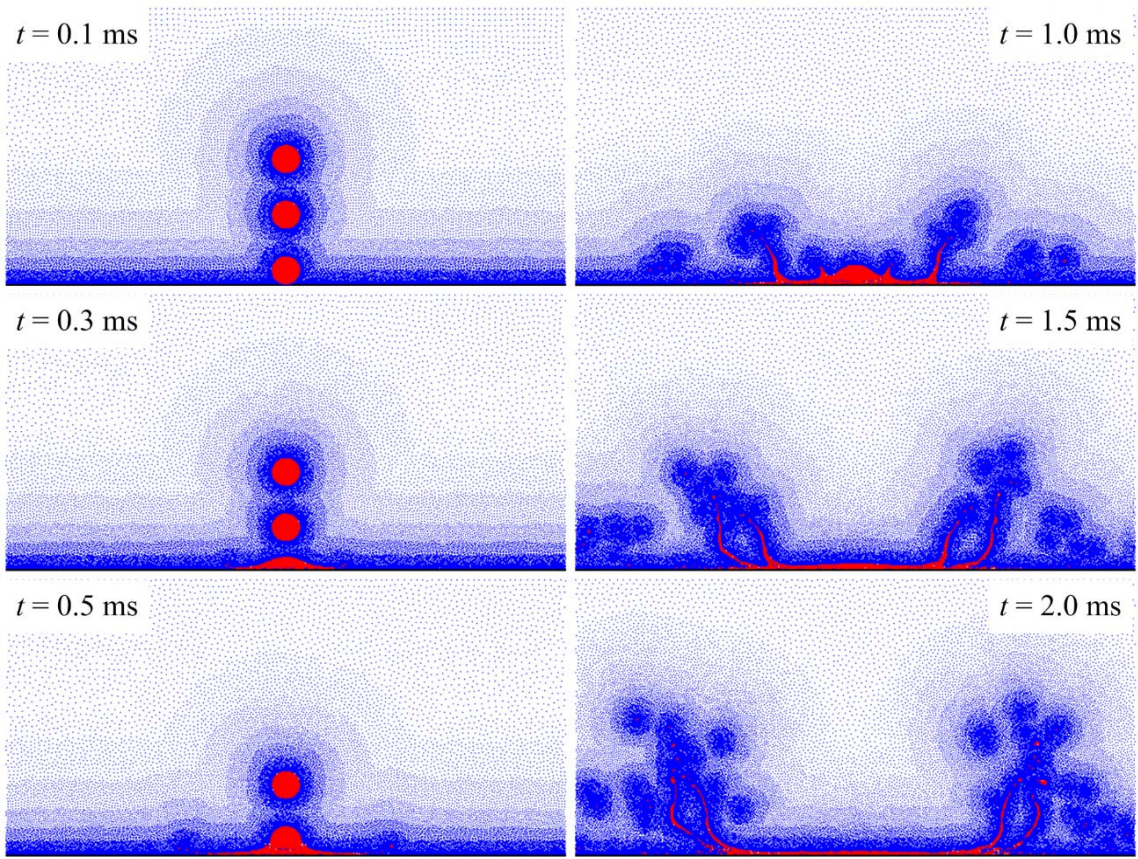

Fig. 13. Simulation results of drop impact on a dry surface using adaptive resolution $C_r$ = 1.1

Fig. 14 compares the numerical results of three consecutive drops impinging on a dry surface. Different resolution scenarios are shown. If uniform resolution Level 3 is regarded as the baseline, all of the adaptive resolutions perform well while using much less particles, as shown in Fig. 15.

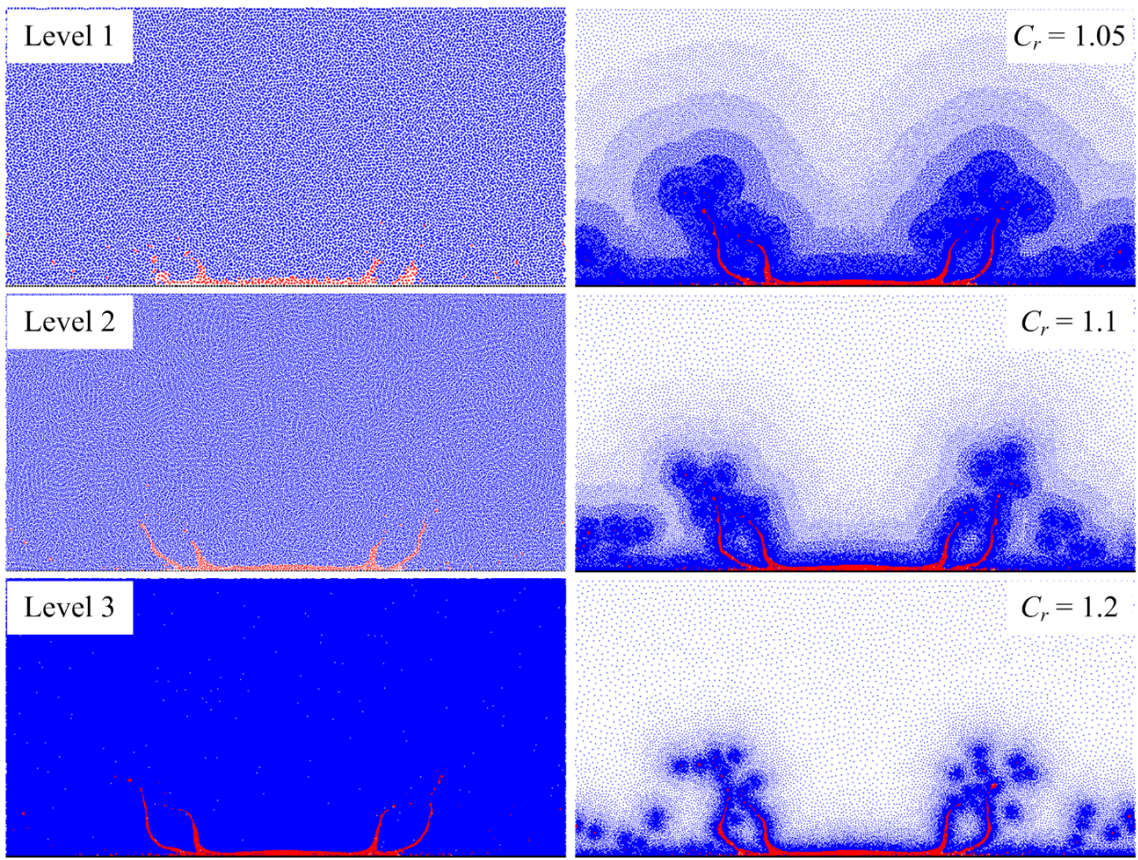

Fig. 14. Comparison of simulation results at $t$ = 1.5 using different uniform resolutions (left) and adaptive resolutions (right)

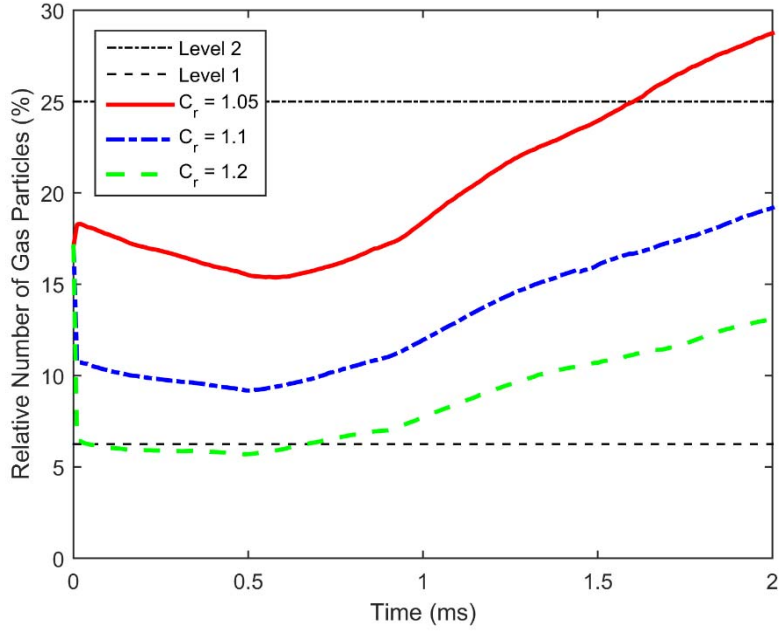

Fig. 15. Relative number of gas particles as a function of time for different resolution scenarios. The relative number of uniform resolution Level 3 is 100%.

Table 3 shows the computational time of the simulations using different resolutions. For uniform resolutions Level 1 to Level 3, the CPU time increases about 8 times when the particle spacing decreases 2 times, which is the same as that in Section 4.1. Comparing with uniform resolution Level 3, the simulations of adaptive resolution $C_r$ = 0.05, 0.1, and 0.2 are about 3.3, 4.7, and 6.7 times faster, respectively.

Table 3. Comparison of CPU time of different resolutions for drop impact on a dry surface.

| Resolution | Particle spacing (mm) | CPU time (minute) |
|---|---|---|
| Level 1 | 0.1 | 9 |
| Level 2 | 0.05 | 76 |
| Level 3 | 0.025 | 607 |
| $C_r$ = 0.05 | 0.025 ~ 0.2 | 186 |
| $C_r$ = 0.1 | 0.025 ~ 0.2 | 129 |
| $C_r$ = 0.2 | 0.025 ~ 0.2 | 90 |

**4.3 Drop impact on a deep pool**

In the above two examples, adaptive resolution was used only in the gas phase to resolve the interface. The next example is to simulate drop impact on a deep pool in which adaptive resolution is used in both the gas and liquid phases. The computational setting is similar to Fig. 6. The pool depth $H_f$ is 10$D$. The distance between the drop and the film surface $H_d$ is $D/2$. The computational domain is 20$D$ in width and 20$D$ in height.

Numerical results are shown in Fig. 16 for different computational scenarios. Different values of adaptive number $C_r$ are used for the gas ($C_r = 1.1$) and liquid ($C_r = 1.05$) phases. It can be seen that the refinement regions have different widths for different phases, demonstrating the flexibility of the present method. As shown in Fig. 16, when the drop impacts the deep pool, a crown is formed, similar to the case of drop impact on a film. At the same time, a cavity is formed in the pool, and the size of the crown-cavity structure increases with time.

Fig. 17 compares the numerical results at $t = 2$ ms using uniform resolution and adaptive resolution. The results using adaptive resolution agree well with those using the finest uniform resolution (Level 3), but the adaptive resolution uses much less particles. As shown in Figs. 18 and 19, the numbers of both gas and liquid particles used in adaptive resolution are very close to the number of particles used in the coarsest uniform resolution (Level 1). However, simulations using adaptive resolution can capture detailed interface evolution and the dynamics of the impact.

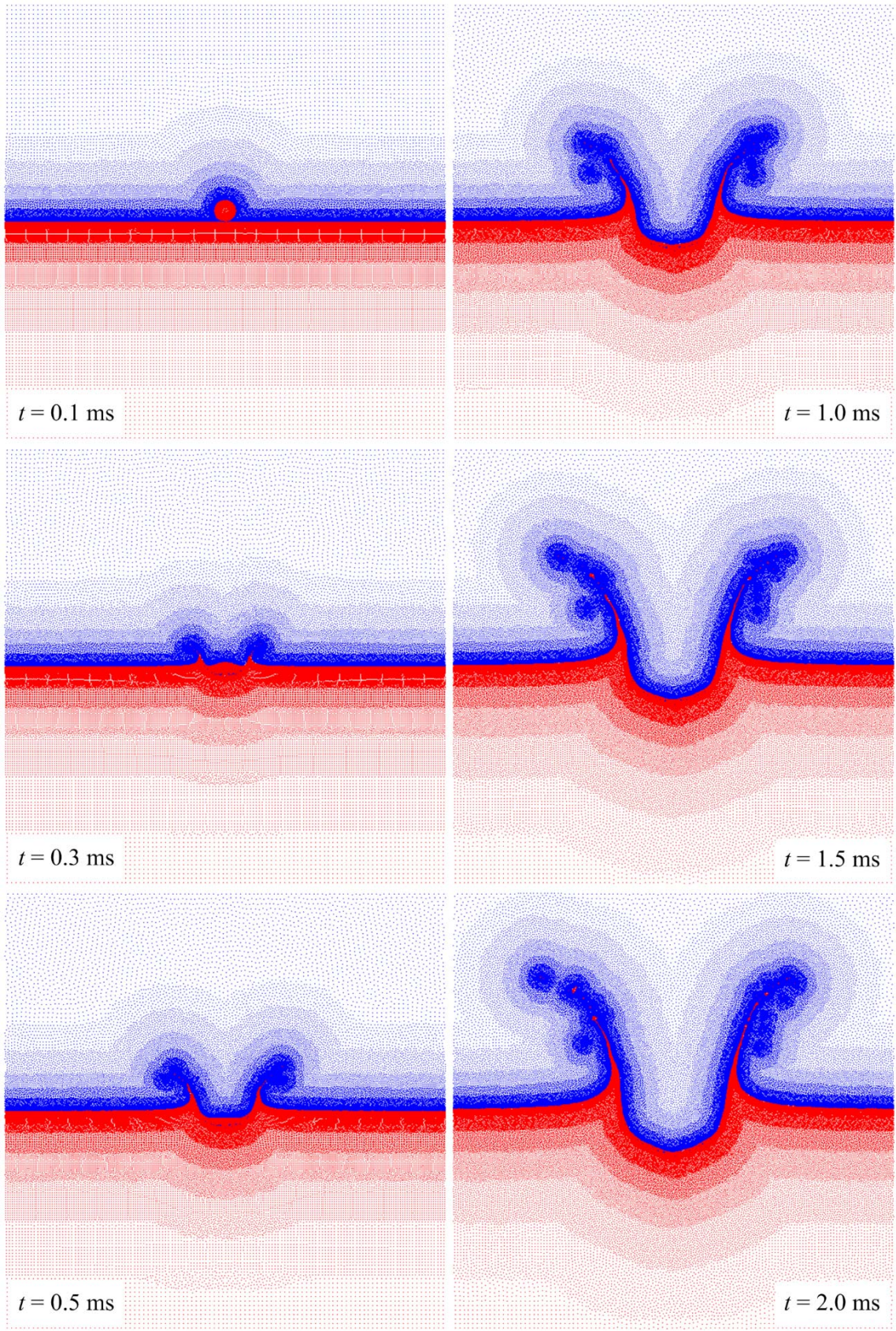

Fig. 16. Simulation results of drop impact on a deep pool with adaptive resolution $C_r$ = 1.1 for the gas phase and $C_r$ = 1.05 for the liquid phase.

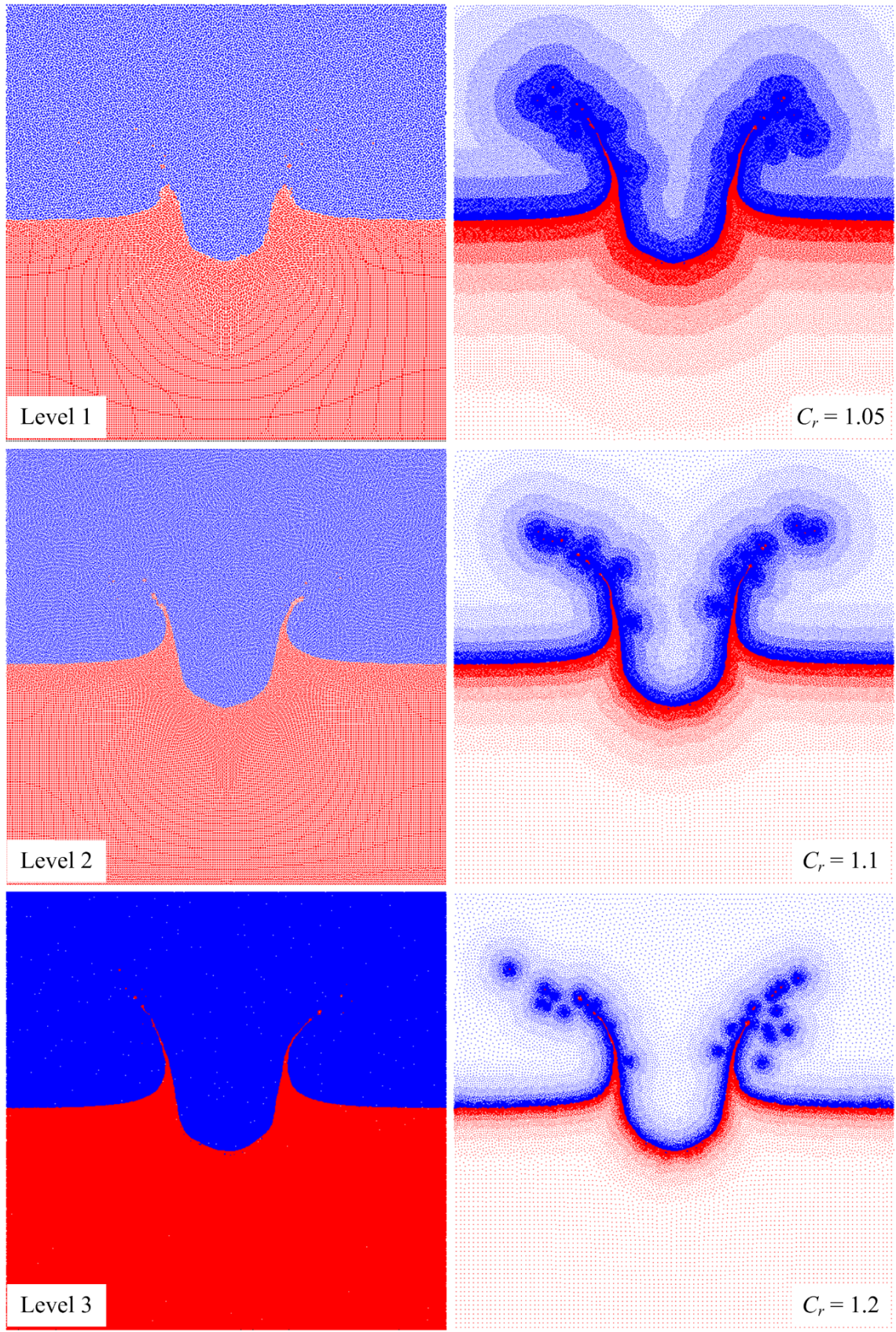

Fig. 17. Comparison of simulation results at $t$ = 2 ms between uniform resolution (left) and adaptive resolutions (right)

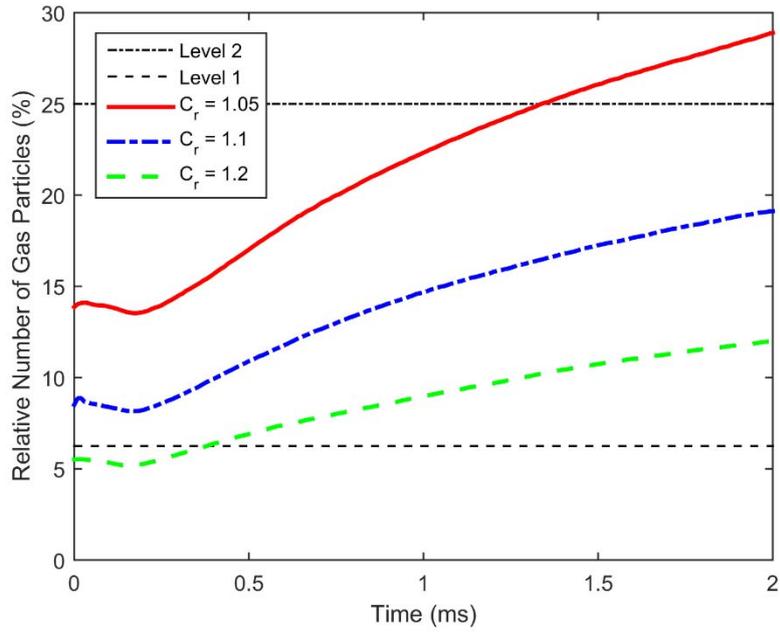

Fig. 18. Relative number of gas particles as a function of time for drop impact on a pool. The relative number of uniform resolution Level 3 is 100%.

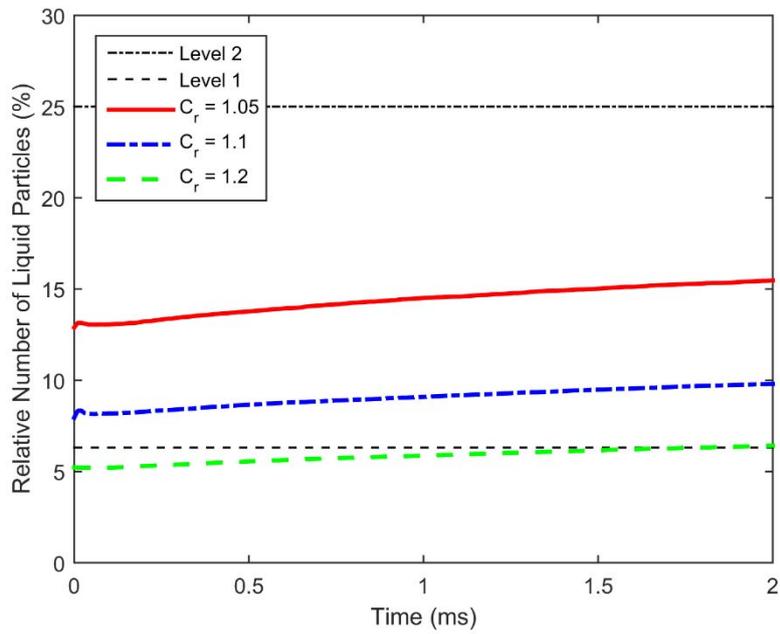

Fig. 19. Relative number of liquid particles as a function of time for drop impact on a pool. The relative number of uniform resolution Level 3 is 100%.

Table 4 shows the computational time of the simulations using different resolutions. The simulations of uniform resolutions are 8 times faster when the particle spacing increases 2 times. The simulations of adaptive resolution $C_r = 0.05$, 0.1, and 0.2 are about 4.4, 6.6, and 10.2 times faster than the uniform resolution Level 3, respectively.

Table 4. Comparison of CPU time of different resolutions for drop impact on a deep pool.

| Resolution | Particle spacing (mm) | CPU time (minute) |
|---|---|---|
| Level 1 | 0.1 | 18 |
| Level 2 | 0.05 | 143 |
| Level 3 | 0.025 | 1148 |
| Cr = 0.05 | 0.025 ~ 0.2 | 259 |
| Cr = 0.1 | 0.025 ~ 0.2 | 174 |
| Cr = 0.2 | 0.025 ~ 0.2 | 113 |

**4.4 Water entry of a horizontal cylinder**

When a solid object impacts on a deep pool, the outcome will be different from that of drop impact. The diameter of the cylinder is 0.11 m, and the initial velocity is 2.955 m/s. The initial water depth is 0.5 m. The particle spacing is in the range of 1 mm to 16 mm. Adaptive resolution with $C_r = 1.2$ is used in the simulation. The simulation results are shown in Fig. 20, comparing with the experimental results and the simulation using constrained interpolation profile (CIP) method [24]. It can be seen that the results from the adaptive resolution SPH agree with the experimental results and are better than the CIP results, because the CIP method does not predict the splashing. Fig. 21 shows the penetration depth of the cylinder in water. The results from three different methods reach a good agreement.

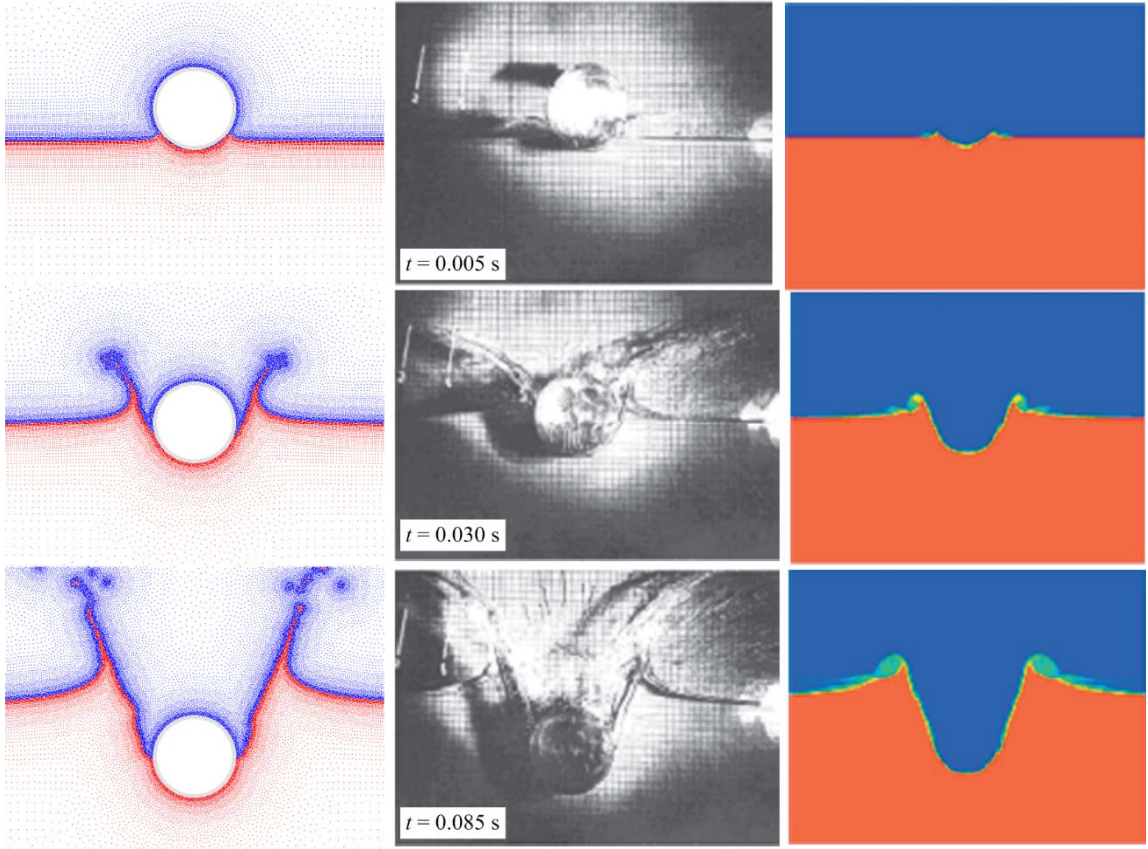

Fig. 20. Comparison of results from adaptive resolution SPH (left), experiments (middle) and CIP (right) [24].

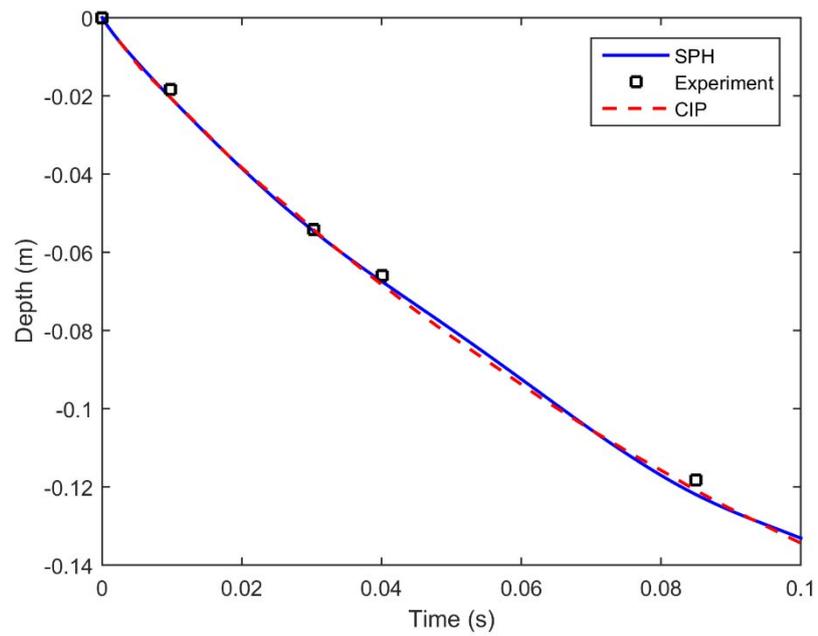

Fig. 21. Penetration depth of cylinder in water.

**4.5 Dam break flow**

Fig. 22 shows the computational setting of dam break. On the left side of the gate is deep water with depth of $H_1 = 0.15$ m and width of $B_1 = 0.38$ m, while on the right side of the gate is a shallow water with depth of $H_2 = 0.018$ m and width of $B_2 = 0.76$ m. The gate is closed initially, then it moves above at a constant velocity $U_0 = 1$ m/s. Due to the gravitational force, the water on the left side will flow to the right side. Adaptive resolution with $C_r = 1.2$ is used in the simulation. The particle spacing varies from 1 mm to 8 mm. Fig. 23 compares the numerical results from adaptive resolution SPH and the experimental results from literature [25]. The "mushroom-like" jet was formed at the first stage, then some air bubbles were trapped by the water waves.

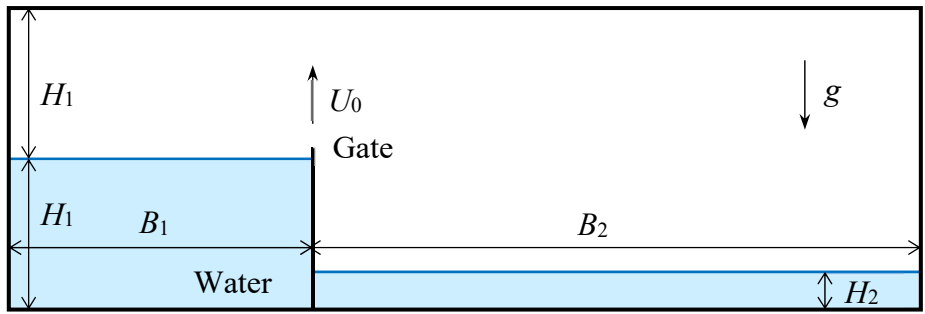

Fig. 22. Computational setting of dam break.

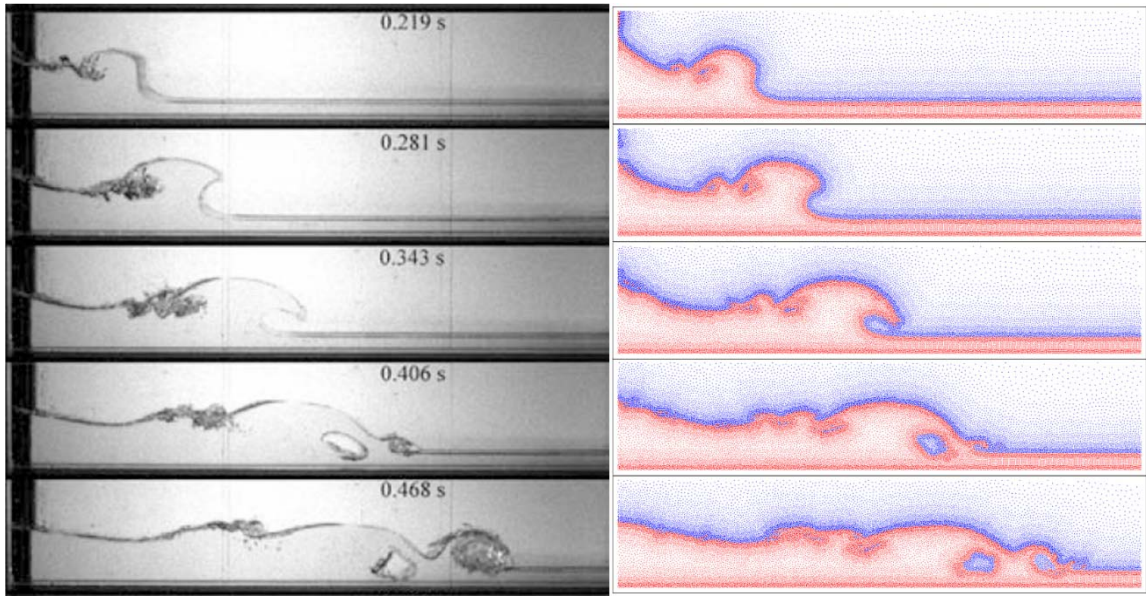

Fig. 23. Comparison between experimental (left) [25] and adaptive resolution SPH (right) results.

## 5. Conclusions and discussions

An adaptive resolution method was developed for SPH simulation of multiphase flows with an application in drop-wall impact. The particles near the interface are refined via particle splitting, while the particles far away from the interface are coarsened via particle merging. A reference particle spacing is defined for each particle such that the resolution can be changed adaptively, following the movement of the interface. The reference particle spacing is a function of the distance between the particle and the interface. As the particle spacing is not uniform, a variable smoothing length is used with the present SPH method. To reduce the numerical errors, the variable smoothing length is further improved by considering the average smoothing length of the neighboring particles.

The present adaptive resolution method was applied to simulate three different drop impact problems, (namely, drop impact on a wet surface, drop impact on a dry surface, and drop impact on a pool), one water entry problem, and one dam break problem. For the drop impact problems, different resolutions are used for simulations. The results show that the present adaptive resolution gives the same results as those obtained using the finest uniform resolution. However, the adaptive method uses much less SPH particles and is significantly more computational efficient. For the water entry and dam break problems, the adaptive resolution SPH are compared with the experiments and CIP method from literature, and they reach a good agreement.

**Acknowledgements**

The authors acknowledge the financial support by Ford Research and Innovation Center.